\documentclass[12pt]{article}
\usepackage{a4wide,epsfig,psfrag,amsmath,amssymb,cite,scalefnt}
\usepackage{color}

\graphicspath{ {figs/} }

\makeatletter
\renewcommand\@makefntext[1]{\leftskip=1em\hskip-1em\@makefnmark#1}
\makeatother

\newcommand{\alsfivepi}{a_s}
\newcommand{\als}{\alpha_s}

\newcommand{\Oonep}{{\cal O}_1^\prime}
\newcommand{\Otwop}{{\cal O}_2^\prime}

\newcommand{\LH}{L_H}
\newcommand{\LHtwo}{L^{\;2}_H}
\newcommand{\LHthr}{L^{\;3}_H}
\newcommand{\LT}{L_t}
\newcommand{\LTtwo}{L^{\;2}_t}
\newcommand{\LTthr}{L^{\;3}_t}

\newcommand{\MH}{M_H}
\newcommand{\MHtwo}{M^{\;2}_H}
\newcommand{\MHthr}{M^{\;3}_H}
\newcommand{\MHfou}{M^{\;4}_H}

\newcommand{\mb}{m_b}
\newcommand{\mbtwo}{m^{\;2}_b}
\newcommand{\mbmh}{\left(\frac{\mbtwo}{\MHtwo}\right)}
\newcommand{\Mt}{M_t}
\newcommand{\Mttwo}{M^{\;2}_t}

\newcommand{\cf}{C_F}
\newcommand{\cftwo}{C^{\,2}_F}
\newcommand{\cfthr}{C^{\,3}_F}
\newcommand{\ca}{C_A}
\newcommand{\catwo}{C^{\,2}_A}
\newcommand{\nl}{n_l}
\newcommand{\nltwo}{n^{\,2}_l}

\newcommand{\zetatwo}{\zeta_2}
\newcommand{\zetathr}{\zeta_3}
\newcommand{\zetafou}{\zeta_4}
\newcommand{\zetafiv}{\zeta_5}
\newcommand{\lifourhalf}{\mbox{Li}_4\left(\frct{1}{2}\right)}
\newcommand{\logtwo}{\ln{2}}
\newcommand{\logtwotwo}{(\ln{2})^2}
\newcommand{\logtwofou}{(\ln{2})^4}

\newcommand{\frct}[2]{\mbox{\Large{$\frac{#1}{#2}$}}}

\allowdisplaybreaks

\begin{document}

\title{\vskip-3cm{\baselineskip14pt
    \begin{flushleft}
      \normalsize TTP17-010
  \end{flushleft}}
  \vskip1.5cm 
  Completing the hadronic Higgs boson decay\\ at order $\als^4$
  \\[1em] }

\author{
  Joshua Davies,
  Matthias Steinhauser,
  David Wellmann
  \\[1em]
  {\small\it Institut f{\"u}r Theoretische Teilchenphysik}\\
  {\small\it Karlsruhe Institute of Technology (KIT)}\\
  {\small\it 76128 Karlsruhe, Germany}  
}
  
\date{}

\maketitle

\thispagestyle{empty}

\begin{abstract}

We compute four-loop corrections to the hadronic decay of the Standard
Model Higgs boson which are induced by effective couplings to bottom
quarks and gluons, mediated by the top quark. 
Our numerical results are comparable in size to
the purely massless contributions which have been known for a few
years. The results presented in this paper complete the order
$\als^4$ corrections to the hadronic Higgs boson decay.

%

\end{abstract}

\thispagestyle{empty}


\newpage


\section{Introduction}

In particle physics, one of the most important tasks in the coming years is
the precise measurement of the couplings of the Higgs boson to fermions and
bosons.  An important ingredient in this context is the decay rate of the
Higgs boson into bottom quarks, which has the by far largest branching
ratio. Together with the decay rate into gluons it constitutes almost 70\%
of the hadronic decay width and it thus has a major influence on
all Higgs boson branching ratios.

One-loop\footnote{In the following we count the number of loops needed for the
  virtual corrections} QCD corrections to $\Gamma(H\to b\bar{b})$ have been
known for a long time, including the full bottom quark--mass
dependence~\cite{Braaten:1980yq}.  The massless approximation\footnote{Here
  ``massless'' refers to the bottom quark mass in the propagators; the bottom
  quark Yukawa coupling remains non-zero.} at order $\als^2$ has been computed
in Ref.~\cite{Gorishnii:1990zu} and the full bottom quark--mass dependence is known
from Ref.~\cite{Chetyrkin:1997mb,Harlander:1997xa,Chetyrkin:1998ix}. Three-
and four-loop corrections, of order $\als^3$ and $\als^4$, have been computed
in the massless limit in
Refs.~\cite{Chetyrkin:1996sr,Chetyrkin:1997vj,Baikov:2005rw}.  A summary of
further corrections, including top quark--mass-suppressed terms and
electroweak effects can be found in recent review
articles~\cite{deFlorian:2016spz,Spira:2016ztx}
(see also the program {\tt HDECAY}~\cite{Djouadi:1997yw}).

The main aim of this paper is to complete the corrections of order
$\als^4$ to the total decay rate of the Higgs boson into hadrons.  In
Ref.~\cite{Baikov:2005rw} only the contribution involving the bottom quark Yukawa
coupling was considered. We
compute the contributions induced by effective Higgs--bottom quark and Higgs--gluon couplings. 
The corresponding three-loop calculation, which was performed in
Ref.~\cite{Chetyrkin:1997vj}, produces a similarly-sized contribution to the $\als^3$ coefficient as that of the purely massless contribution. It is therefore
necessary also to evaluate the top quark--induced contributions at order $\als^4$.

For the calculation performed in this paper the relevant part the Standard Model (SM)
Lagrange density is given by the Yukawa terms supplemented by the
strong interaction terms. For the production and decay of the SM Higgs boson it
turns out that the effective theory in which the top quark is integrated
out provides a good approximation to the full theory. This leads to the following effective
Lagrangian~\cite{Inami:1982xt,Chetyrkin:1996wr,Chetyrkin:1996ke}\footnote{We follow the notation of Ref.~\cite{Chetyrkin:1997vj}.}
\begin{eqnarray}
  {\cal L}_{\rm eff} &=& -\frac{H^0}{v^0}
  \left( C_1 [\Oonep] + C_2 [\Otwop] \right)
  + {\cal L}_{\rm QCD}^\prime
  \,,
  \label{eq::leff}
\end{eqnarray}
where the primed quantities are defined in the five-flavour theory.
$H^0$ and $v^0$ are the bare Higgs boson
field and vacuum expectation value which can be identified with their
renormalized counterparts if, as in this paper, electroweak effects are neglected.
In Eq.~(\ref{eq::leff}) all dependence on the top quark is
contained in the coefficient functions (or effective couplings) $C_1$
and $C_2$. $[\Oonep]$ and $[\Otwop]$ are
renormalized effective operators constructed from the light degrees of
freedom. Their bare versions read
\begin{eqnarray}
  \Oonep &=& \left(G_{a,\mu\nu}^{0\prime}\right)^2
  \,,\nonumber\\
  \Otwop &=& \mb^{0\prime} \bar{b}^{0\prime} b^{0\prime}\,,
\end{eqnarray}
where $G_{a,\mu\nu}^{0\prime}$ is the bare gluon field strength tensor
and $\bar{b}^{0\prime}$ is the bare bottom quark field.

Further corrections to ${\cal L}_{\rm eff}$ are suppressed by the inverse top
quark mass, contributing terms of order $\MHtwo/\Mttwo$ to the decay
rate. These terms are available to order
$\als^3$~\cite{Chetyrkin:1995pd,Larin:1995sq,Schreck:2007um} and are known to
be small. For example, at order $\als^2$ the $\MHtwo/\Mttwo$ term changes the
coefficient by less than 1\% and thus induces a correction which is of the
same order of magnitude as non-suppressed contributions of order $\als^4$.  We
also restrict ourselves to the leading $\mbtwo$ term and neglect higher powers
in the bottom quark mass which are numerically even smaller than the $1/\Mt$
terms.

\begin{figure}[t]
  \centering
  \includegraphics[width=\textwidth]{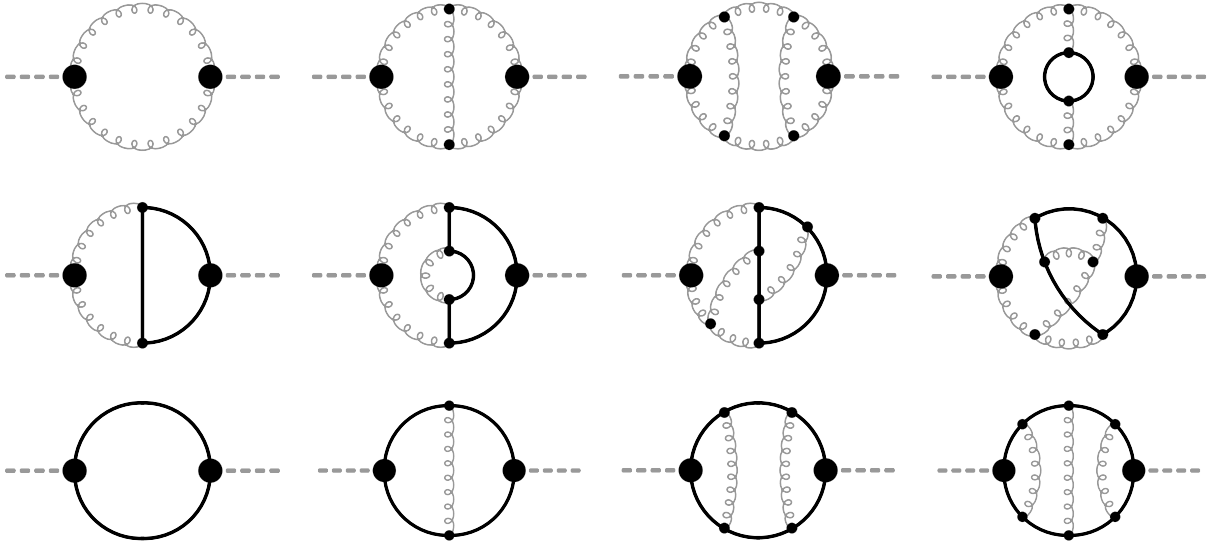}
  \caption{\label{fig::diag}Sample Feynman diagrams contributing to
    $\Pi_{11}$, $\Pi_{12}$ and $\Pi_{22}$. The curly and straight
    lines represent gluons and quarks, respectively. The blobs stand
    for the effective operators $\Oonep$ and $\Otwop$.}
\end{figure}

On the basis of the Lagrange density of Eq.~(\ref{eq::leff}) we define
correlators formed by the operators $\Oonep$ and $\Otwop$,
\begin{eqnarray}
  \Pi_{ij}(q^2) &=& i\int {\rm d} x e^{iqx}
  \langle 0 | T[{\cal O}_i^\prime,{\cal O}_j^\prime ] |0 \rangle.
  \nonumber\\
\end{eqnarray}
Sample Feynman diagrams contributing to $\Pi_{11}$,
$\Pi_{12}$ and $\Pi_{22}$ are shown in Fig.~\ref{fig::diag}.

Using the optical theorem, the total decay rate can be obtained from the
imaginary part of $\Pi_{ij}$. In this context it is convenient to
introduce the quantities
\begin{eqnarray}
  \Delta_{ii} &=& K_{ii} \,\mbox{Im}\left[ \Pi_{ii}(\MHtwo) \right]
  \,,
  \nonumber\\
  \Delta_{12} &=& K_{12} \,\mbox{Im}\left[ \Pi_{12}(\MHtwo) + \Pi_{21}(\MHtwo) \right]
  \,,
\end{eqnarray}
with $1/K_{11} = 32\pi \MHfou$ and $1/K_{12} = 1/K_{22} = 6\pi \MHtwo \mbtwo$.
Note that $\Pi_{12}(\MHtwo) = \Pi_{21}(\MHtwo)$.
The total decay width is then given by
\begin{eqnarray}
  \Gamma(H\to\mbox{hadrons}) &=& 
  A_{b\bar{b}} \left[
    \left(C_2\right)^2 \left(1+\Delta_{22}\right) 
    + C_1 C_2 \Delta_{12} 
    \right]
  +A_{gg} \left(C_1\right)^2 \Delta_{11} 
  \,,
  \label{eq::Gam}
\end{eqnarray}
where
\begin{eqnarray}
  A_{b\bar{b}} &=& \frac{3 G_F \MH \mbtwo(\mu)}{4\pi\sqrt{2}}\,,\nonumber\\
  A_{gg} &=& \frac{4 G_F \MHthr}{\pi\sqrt{2}}\,.
  \label{eq::AbbAgg}
\end{eqnarray}
Note that for clarity, we restrict ourselves in Eq.~(\ref{eq::Gam}) to the QCD
corrections that we compute in this paper; we neglect both electroweak effects and
power corrections suppressed by $\MHtwo/\Mttwo$.
Furthermore, we concentrate on the decay of the Higgs boson only to
bottom quarks and to gluons. The results can easily be extended to include the decay
to additional light quark flavours, if necessary.
A more complete formula can be found in Eq.~(10) of Ref.~\cite{Chetyrkin:1997vj}. 
Note that in Eq.~(\ref{eq::AbbAgg}), $\mb(\mu)$ refers to the $\overline{\rm MS}$
bottom quark mass evaluated at the renormalization scale $\mu$.

In Ref.~\cite{Baikov:2005rw} $\Pi_{22}$ has been computed to five-loop order,
yielding order $\als^4$ corrections to the Higgs boson decay. For these
corrections we have that $C_2=1$ and therefore refer to them in the following as
``massless contributions'', despite the fact that there is an overall factor of
$\mbtwo$ from the bottom quark Yukawa coupling.

The leading-order term of $\Pi_{11}$ describes the decay of the Higgs boson
into gluons. Starting from next-to-leading order (two loops) the gluonic and fermionic decay
cannot be separated in the approach based on the optical theorem, 
since there are diagrams containing both purely
gluonic cuts and cuts involving both gluons and quark--antiquark pairs.

The main result of this paper is the extension of~\cite{Chetyrkin:1997vj}. We
compute the four-loop correction to $\Pi_{12}$ which contributes to the hadronic
Higgs boson decay at order $\als^4$, along with the five-loop calculation of
Ref.~\cite{Baikov:2005rw}. This is because the leading term of $C_1$ contains
a factor $\als$.

Note that $\Pi_{22}$ has an overall prefactor $\mbtwo$, which comes from the two
operators $\Otwop$. $\Pi_{12}$ is also proportional to $\mbtwo$; one
factor arises from $\Otwop$ the other from the trace of the bottom
quark loop. In the limit $\mb\to0$ the correlator $\Pi_{11}$ has a non-zero
contribution. Terms proportional to $\mbtwo$ appear for the first time at two-loop
order, due to the presence of closed bottom quark loops.
We compute such terms up to three loops, which give
rise to order $\als^4$ corrections to the Higgs boson decay. We want
to remark that the $\mb$-independent terms of $\Pi_{11}$ have been computed to four-loop order
in Ref.\cite{Baikov:2006ch} leading to corrections of order $\als^5$ to
the hadronic Higgs boson decay.

In the next section we provide several technical details of our
calculation. In particular, we discuss the computation of the four-loop
integrals and explain the operator mixing and renormalization. We additionally
provide explicit expressions for the effective couplings $C_1$ and $C_2$.
We present analytic results in Section~\ref{sec::res} and discuss
the numerical impact of our new corrections. Our conclusions are given in
Section~\ref{sec::concl}.


\section{\label{sec::calc}Calculation}

For the calculation of the Feynman diagrams we use a well-tested automated
setup which uses {\tt qgraf}~\cite{Nogueira:1991ex} for the generation of the
Feynman amplitudes, and {\tt q2e} and {\tt
  exp}~\cite{Harlander:1997zb,Seidensticker:1999bb,q2eexp} for the mapping to
one of eleven pre-defined four-loop integral families. The Dirac algebra is
performed with {\tt FORM}~\cite{Kuipers:2012rf}, which also re-writes the
amplitude of each diagram as a linear combination of scalar integrals. Next we generate,
using {\tt FIRE 5.1}~\cite{Smirnov:2014hma,FIRE}, tables for the
reduction of the integrals of all eleven families to master integrals.  We then
apply {\tt tsort}~\cite{Smirnov:2013dia}, in the form of the {\tt FIRE}
command {\tt FindRules}, to minimize the number of master integrals among all
eleven families and end up with 28 four-loop master integrals, which have been
computed in Refs.~\cite{Baikov:2010hf,Smirnov:2010hd,Lee:2011jt}.

We have re-computed the one-, two- and three-loop corrections to all
correlators using both the setup described above and, as a cross check of our
approach, {\tt MINCER}~\cite{Larin:1991fz}. Both calculations produce
identical results, which agree with the literature. As a further check we have
performed our calculations using a generic gauge parameter $\xi$. Our
four-loop expressions have been expanded to linear order in $\xi$ which drops
out after reducing the master integrals to a minimal set.

We have used this method to compute the four-loop corrections
to $\Pi_{12}$ and the three-loop corrections to $\Pi_{11}$ which, after taking
the imaginary part, lead to the bare quantities $\Delta_{12}^0$ and
$\Delta_{11}^0$. At this point we perform the renormalization of the
strong coupling constant and the quark mass in the $\overline{\rm MS}$
scheme where the renormalization constants are introduced via
\begin{eqnarray}
  \als^0 &=& Z_{\als} \als\,,\nonumber\\
  \mb^0 &=& Z_m \mb\,.
\end{eqnarray}
$Z_{\als}$ and $Z_m$ are required to third order in $\als$
and can be found in, e.g., Ref.~\cite{Chetyrkin:2004mf}.

Afterwards, we have to take into account that the operators
$\Oonep$ and $\Otwop$ mix under
renormalization according to~\cite{Spiridonov:1984br,Chetyrkin:1997vj}
\begin{eqnarray}
  [\Oonep] &=& Z_{11} \Oonep + Z_{12} \Otwop 
  \,,\nonumber\\
  \mbox{}[\Otwop] &=& \Otwop
  \,.
\end{eqnarray}
The renormalization constants $Z_{11}$ and $Z_{12}$
are obtained from $Z_{\als}$ and $Z_m$ as follows,
\begin{eqnarray}
  Z_{11} &=& 1+\frac{\als\partial}{\partial\als}\log Z_{\als}
  \,,\nonumber\\
  Z_{12} &=& - 4 \frac{\als\partial}{\partial\als}\log Z_m
  \,.
\end{eqnarray}
In terms of these renormalization constants, the renormalized correlators $\Delta_{ij}$ are given by
\begin{eqnarray}
  \Delta_{11} &=& (Z_{11})^2 \Delta_{11}^0 + 2 Z_{11}Z_{12}
  \Delta_{12}^0 + (Z_{12})^2 \Delta_{22}^0 
  \,,\nonumber\\
  \Delta_{12} &=& Z_{11} \Delta_{12}^0 + Z_{12} \Delta_{22}^0
  \,,\nonumber\\
  \Delta_{22} &=& \Delta_{22}^0
  \,.
\end{eqnarray}
Note that the contributions of $\Delta_{12}^0$ and $\Delta_{22}^0$ are
proportional to $\mbtwo$ whereas $\Delta_{11}^0$ contains both
$\mbtwo$ and $\mb$-independent terms.
Since $Z_{12}\propto\als$ only the $(n-1)$-loop terms of $\Delta_{12}^0$
and the $(n-2)$-loop terms of $\Delta_{22}^0$ enter the $n$-loop
renormalization of $\Delta_{11}$.
Similarly, the $n$-loop renormalization of $\Delta_{12}$ requires the
$(n-1)$-loop terms of $\Delta_{22}^0$.

For completeness we also provide explicit expressions for the effective
couplings $C_1$ and $C_2$, which are available in the literature up to fifth
order~\cite{Schroder:2005hy,Chetyrkin:2005ia,Liu:2015fxa}.  It is convenient
to parametrize the perturbative expansion in terms of
\begin{eqnarray}
	\alsfivepi &\equiv& \frac{\als^{(5)}(\mu)}{\pi}\,,
\end{eqnarray}
where the superscript indicates the number of active quark flavours used for
the running, and the on-shell top quark mass.  To obtain corrections of order
$\alsfivepi^4$ to the decay rate, $C_1$ is needed to third order and $C_2$ to
fourth order.  In the following we present $C_1$ to order $\alsfivepi^4$ since
we include $\alsfivepi^5$ corrections when evaluating the decay rate
numerically.  The analytic results read
\begin{align}
C_1 = \: &
- \alsfivepi \, \frct{1}{12}
- \alsfivepi^2\, \frct{11}{48}
- \alsfivepi^3 \, \bigg[
	\frct{2777}{3456}
	+ \frct{19}{192} \, \LT
	- \nl \, \Big(
		\frct{67}{1152}
		- \frct{1}{36} \, \LT
	\Big)
\bigg]
\nonumber\\*&
+ \alsfivepi^4 \, \bigg[
	\frct{2761331}{497664}
	- \frct{897943}{110592}\,\zetathr
	- \frct{2417}{3456}\,\LT
	- \frct{209}{768}\,\LTtwo
\nonumber\\*&\hphantom{{}+ \alsfivepi^4 \, \bigg[}
	- \nl \, \Big(
		\frct{58723}{248832}
		- \frct{110779}{165888}\,\zetathr
		+ \frct{91}{648}\,\LT
		+ \frct{23}{384}\,\LTtwo
	\Big)
\nonumber\\*&\hphantom{{}+ \alsfivepi^4 \, \bigg[}
	+ \nltwo \, \Big(
		\frct{6865}{373248}
		- \frct{77}{20736}\,\LT
		+ \frct{1}{216}\,\LTtwo
	\Big)
\bigg]	
+ {\cal O}\left( \alsfivepi^5 \right),
\\*
\approx \: &
- 0.08333 \, \alsfivepi
- 0.2292 \, \alsfivepi^2
- \alsfivepi^3 \, \Big[
	0.7391
	- 0.07624 \, \nl
\Big]
\nonumber\\*&
- \alsfivepi^4 \, \Big[
	3.8715
	- 0.6328 \, \nl
	- 0.02277 \, \nltwo
\Big]
+ {\cal O}\left( \alsfivepi^5 \right),
\\*
C_2 = \: &
1
+ \alsfivepi^2 \, \bigg[
	\frct{5}{18}
	- \frct{1}{3} \, \LT
	\bigg]
+ \alsfivepi^3 \, \bigg[
	- \frct{841}{1296}
	+ \frct{5}{3} \, \zetathr
	- \frct{79}{36} \, \LT
	- \frct{11}{12} \, \LTtwo
	+ \nl \, \Big(
		\frct{53}{216}
		+ \frct{1}{18} \, \LTtwo
	\Big)
\bigg]
\nonumber\\&\hphantom{1}
+ \alsfivepi^4 \, \bigg[
	\frct{609215}{186624}
	- \frct{4}{3}\,\zetatwo
	+ \frct{374797}{13824}\,\zetathr
	- \frct{4123}{144}\,\zetafou
	- \frct{575}{36}\,\zetafiv
	+ \frct{62}{9}\,\lifourhalf
	- \frct{4}{9}\,\logtwo\,\zetatwo
\nonumber\\&\hphantom{1 + \alsfivepi^4 \, \bigg[}
	- \frct{31}{18}\,\logtwotwo\,\zetatwo
	+ \frct{31}{108}\,\logtwofou
	- \Big[\frct{4645}{144}-\frct{55}{4}\,\zetathr\Big]\,\LT
	- \frct{91}{8}\,\LTtwo
	- \frct{121}{48}\,\LTthr
\nonumber\\&\hphantom{1 + \alsfivepi^4 \, \bigg[}
	+ \nl \, \Big(
		- \frct{11557}{15552}
		+ \frct{2}{9}\,\zetatwo
		- \frct{221}{288}\,\zetathr
		+ \frct{163}{72}\,\zetafou
		- \frct{4}{9}\,\lifourhalf
		+ \frct{1}{9}\,\logtwotwo\,\zetatwo
\nonumber\\&\hphantom{1 + \alsfivepi^4 \, \bigg[ + \nl \, \Big(}
		- \frct{1}{54}\,\logtwofou
		+ \frct{9535}{2592}\,\LT
		+ \frct{109}{144}\,\LTtwo
		+ \frct{11}{36}\,\LTthr
	\Big)
\nonumber\\&\hphantom{1 + \alsfivepi^4 \, \bigg[}
	+ \nltwo \, \Big(
		\frct{3401}{23328}
		- \frct{7}{54}\,\zetathr
		- \frct{31}{324}\,\LT
		- \frct{1}{108}\,\LT^3
	\Big)
\bigg]
+ {\cal O}\left( \alsfivepi^5 \right),
\\
\approx \: &
1
+ 0.494759 \, \alsfivepi^2
+ \alsfivepi^3\, \Big[
	2.3946
	+ 0.2689 \, \nl
\Big]
\nonumber\\&\hphantom{1}
- \alsfivepi^4 \, \Big[
	6.0125
	+1.1543 \, \nl
	-0.05480 \, \nltwo
\Big]
+ {\cal O}\left( \alsfivepi^5 \right),
\end{align}
where $\zeta_n$ is the Riemann Zeta function, $\mbox{Li}_n(z)$ is the
Polylogarithm function and we have defined $\LT = \log(\mu^2/\Mttwo)$. The
numerical expressions are given at the renormalization scale $\mu^2 = \MHtwo$,
for $\nl = 5$ massless flavours running in fermion loops, and for
$\MH=125.09$~GeV and $\Mt=173.21$~GeV~\cite{Olive:2016xmw}.
Since $\mu$ is of the order of the
Higgs boson mass one generates potentially large logarithms which should be
resummed~\cite{Chetyrkin:1996ke}. In practice, however, the numerical effect
is small and we have decided to consider only the fixed-order result here.


\section{\label{sec::res}Results}

We use this section to present our results. The new ingredients of
Eq.~(\ref{eq::Gam}) required to complete the order $\als^4$ corrections
to $\Gamma(H\to\mbox{hadrons})$ are the four-loop corrections to
$\Delta_{12}$ and the bottom mass--dependent three-loop corrections to
$\Delta_{11}$. For convenience we also present the lower-order contributions.
The general expressions in terms of the Casimir invariant colour factors can be
found in Appendix~\ref{app::Del12}.
For the SU(3) case, for which $\ca = 3$ and $\cf = 4/3$, we obtain
\begin{align}
\Delta_{11} = \: &
1
+\alsfivepi^{\hphantom{2}} \,
\bigg[
	\frct{73}{4}
	+ \frct{11}{2}\,\LH
	- \nl \, \Big(
		\frct{7}{6}
		+ \frct{1}{3}\,\LH
	\Big)
\bigg]
\nonumber\\*&\hphantom{1}
+\alsfivepi^2 \, \bigg[
	\frct{37631}{96}
	- \frct{363}{8}\,\zetatwo
	- \frct{495}{8}\,\zetathr
	+ \frct{2817}{16}\,\LH
	+ \frct{363}{16}\,\LHtwo
\nonumber\\&\hphantom{1+\alsfivepi^2 \, \bigg[}
	+ \nl \, \Big(
		- \frct{7189}{144}
		+ \frct{11}{2}\,\zetatwo
		+ \frct{5}{4}\,\zetathr
		- \frct{263}{12}\,\LH
		- \frct{11}{4}\,\LHtwo
	\Big)
\nonumber\\&\hphantom{1+\alsfivepi^2 \, \bigg[}
	+ \nltwo \, \Big(
		\frct{127}{108}
		- \frct{1}{6}\,\zetatwo
		+ \frct{7}{12}\,\LH
		+ \frct{1}{12}\,\LHtwo
	\Big)
\bigg]
\nonumber\\&
+ \mbmh \, \Bigg\{
	6 \, \alsfivepi
	+ \alsfivepi^2 \, \bigg[
		\frct{697}{3}
		- 6\,\zetatwo
		+ 6\,\zetathr
		+ \frct{169}{2}\,\LH
		+ 3\,\LHtwo
		- \nl \, \Big(
			\frct{15}{2}
			+ 3\,\LH
		\Big)
	\bigg]
\Bigg\}
\nonumber\\&
+ {\cal O}\left( \alsfivepi^3 \right)
\\
\approx \: &
1
+ \alsfivepi \, \Big[18.2500 - 1.1667\,\nl \Big]
+ \alsfivepi^2 \, \Big[ 242.9734 - 39.3739\,\nl + 0.9018\,\nltwo \Big]
\nonumber\\&
+ \mbmh \, \Big\{
6 \, \alsfivepi
+ \alsfivepi^2 \, \Big[ 229.6761 - 7.5000\,\nl \Big]
\Big\}
+ {\cal O}\left( \alsfivepi^3 \right)
\\
\intertext{and}
\Delta_{12} =&
+ \alsfivepi^{\hphantom{2}}\,
\bigg[-\frct{92}{3} - 8 \, \LH\bigg]
\nonumber\\&
+ \alsfivepi^2 \, \bigg[
	- \frct{15073}{18}
	+ 76\,\zetatwo
	+ 156 \, \zetathr
	- \frct{1028}{3}\, \LH
	- 38 \, \LHtwo
\nonumber\\*&\hphantom{{} + \alsfivepi^2 \, \bigg[}
	+ \nl \, \Big(
		\frct{283}{9}
		- \frct{8}{3}\,\zetatwo
		- \frct{16}{3}\,\zetathr
		+ \frct{112}{9}\,\LH
		+ \frct{4}{3}\,\LHtwo
	\Big)
\bigg]
\nonumber\\&
+ \alsfivepi^3 \, \bigg[
	- \frct{8957453}{432}
	+ 4150\,\zetatwo
	+ \frct{131389}{18}\,\zetathr
	- 815\,\zetafiv
\nonumber\\*&\hphantom{{} + \alsfivepi^3 \, \bigg[}
	- \Big[\frct{65267}{6} - 855\,\zetatwo - 1755\,\zetathr\Big]\,\LH
	- 2075\,\LHtwo
	- \frct{285}{2}\,\LHthr
\nonumber\\*&\hphantom{{} + \alsfivepi^3 \, \bigg[}
	+ \nl \, \Big(
		\frct{279451}{162}
		- \frct{1003}{3}\,\zetatwo
		- 446\,\zetathr
		+ 10\,\zetafou
		+ \frct{100}{3}\,\zetafiv
\nonumber\\&\hphantom{{} + \alsfivepi^3 \, \bigg[ + \nl \, \Big(}
		+ \Big[\frct{15973}{18} - 68\,\zetatwo - 118\,\zetathr\Big]\,\LH
		+ \frct{1003}{6}\,\LHtwo
		+ \frct{34}{3}\,\LHthr
	\Big)
\nonumber\\&\hphantom{{} + \alsfivepi^3 \, \bigg[}
	+ \nltwo \, \Big(
		- \frct{25627}{972}
		+ \frct{56}{9}\,\zetatwo
		+ \frct{20}{3}\,\zetathr
		- \Big[\frct{407}{27} - \frct{4}{3}\,\zetatwo
			- \frct{8}{3}\,\zetathr\Big]\,\LH
\nonumber\\&\hphantom{+ \alsfivepi^3 \, \bigg[+ \nltwo \, \Big(}
		- \frct{28}{9}\,\LHtwo
		- \frct{2}{9}\,\LHthr
	\Big)
\bigg]
+ {\cal O}\left( \alsfivepi^4 \right)
\\
\approx \: &
-30.6667\,\alsfivepi
+ \alsfivepi^2 \, \Big[ - 524.8530 + 20.6470\,\nl \Big]
\nonumber\\&
+ \alsfivepi^3 \, \Big[ - 5979.1838 + 684.320\,\nl - 8.1164\,\nltwo \Big]
+ {\cal O}\left( \alsfivepi^4 \right),
\end{align}
where $\LH=\log(\mu^2/\MHtwo)$. As above, $\nl$ counts the number of light
quarks running in fermion loops.  For the numerical evaluation we have set
$\mu^2=\MHtwo$ and $\nl=5$.  For both $\Delta_{11}$ and $\Delta_{12}$ we
observe a rapid growth of the coefficients, however, we postpone discussion of
the convergence properties to the decay rate, since $\Delta_{11}$ and
$\Delta_{12}$ do not themselves represent physical quantities.

For the numerical evaluation of the decay rate it is convenient
to cast Eq.~(\ref{eq::Gam}) in the form
\begin{eqnarray}
  \Gamma(H\to\mbox{hadrons}) &=& 
  A_{b\bar{b}} \left( 1 + \Delta_{\rm light} + \Delta_{\rm top} + \Delta_{gg}^{\mb=0} \right)
  \,,
  \label{eq::gamhbb}
\end{eqnarray}
where we have chosen $A_{b\bar{b}}$ as a common prefactor so that we can
easily compare the relative sizes of the individual contributions.
$\Delta_{\rm light}$ contains all corrections obtained for $C_1=0$ and
$C_2=1$. They have already been presented and discussed in
Ref.~\cite{Baikov:2005rw}. $\Delta_{\rm top}$ contains the top quark--induced
corrections obtained from the contributions proportional to $C_1$ and
$(C_2-1)$. For completeness we also list the corrections from $\Delta_{11}$
which have no factor $\mbtwo$. They are collected in
$\Delta_{gg}^{\mb=0}$. Note that these terms have already been computed in
Ref.~\cite{Chetyrkin:1997iv}. For convenience we provide the formulae
which relate the quantities in Eq.~(\ref{eq::gamhbb}) to the ones
in Eq.~(\ref{eq::Gam}):
\begin{eqnarray}
  \Delta_{\rm light} &=& \Delta_{22}\,,\nonumber\\
  \Delta_{\rm top} &=& \left[\left(C_2\right)^2-1\right] \left(1+\Delta_{22}\right) 
  + C_1 C_2 \Delta_{12}
  + \frac{16\MHtwo}{3\mbtwo} \left(C_1\right)^2 \Delta_{11}^{\mbtwo}\,,\nonumber\\ 
  \Delta_{gg}^{\mb=0} &=& \frac{16\MHtwo}{3\mbtwo} \left(C_1\right)^2 \Delta_{11}^{\mb=0} \,.
  \label{eq::Delta}
\end{eqnarray}

For the numerical evaluation we use
$\als^{(5)}(M_Z)=0.1181$~\cite{Olive:2016xmw} and 
$\mb(\mb)=4.163$~GeV~\cite{Chetyrkin:2009fv}
which leads to 
$\mb(\MH) = 2.773$~GeV and $\als^{(5)}(\MH)=0.1127$ 
using {\tt RunDec}~\cite{Chetyrkin:2000yt,Schmidt:2012az} with four-loop accuracy.
Numerical values for $\MH$ and $\Mt$ are already given at the end
of Section~\ref{sec::calc}.
We expand the expressions of Eq.~(\ref{eq::Delta}) in $\alsfivepi$ and obtain
\begin{eqnarray}
  \Delta_{\rm light} &\approx&
  5.6667 \alsfivepi  + 29.1467 \alsfivepi^2 + 41.7576 \alsfivepi^3 - 825.7466 \alsfivepi^4
  \nonumber\\*
  &\approx& 0.2033 + 0.03752 + 0.001929 - 0.001368,
    \label{eq::Delta_light}\\[3mm]
  \Delta_{\rm top} &\approx&
  \alsfivepi^2 \, [2.5556_{12} + 0.9895_{22} ]
  \nonumber\\*&&\mbox{}
  + \alsfivepi^3 \, [0.2222_{11} + 42.1626_{12} + 13.0855_{22}]
  \nonumber\\*&&\mbox{}
  + \alsfivepi^4 \, [8.3399_{11} + 338.9021_{12} + 50.6346_{22}]
  \nonumber\\*
  &\approx&
  0.003290_{12} + 0.001274_{22} 
  \nonumber\\*&&\mbox{}
  + 0.00001026_{11} + 0.001947_{12} + 0.0006043_{22}
  \nonumber\\*&&\mbox{}
  + 0.00001382_{11} + 0.0005616_{12} + 0.00008390_{22},
  \label{eq::Delta_top}
  \\[3mm]
  \Delta_{gg}^{\mb=0} &\approx&
  \frac{\MHtwo}{27\mbtwo}\left(
    \alsfivepi^2 + 17.9167 \alsfivepi^3 + 153.0921 \alsfivepi^4
    + 392.6176 \alsfivepi^5
  \right),
  \nonumber\\*
  &\approx&
  0.09699 + 0.06235 + 0.01911 + 0.001759
  \,,
  \label{eq::Delta_gg}
\end{eqnarray}
where the subscripts in the expression for $\Delta_{\rm top}$ indicate
the origin of each term. For $\Delta_{gg}^{\mb=0}$ we have
included the corrections of order $\alsfivepi^5$ from Ref.~\cite{Baikov:2006ch}.

From Eqs.~(\ref{eq::Delta_light}),~(\ref{eq::Delta_top})
and~(\ref{eq::Delta_gg}) we observe that the $\alsfivepi^2$ term of
$\Delta_{gg}^{\mb=0}$ amounts to almost 50\% of the $\alsfivepi$ term in
$\Delta_{\rm light}$. Furthermore, the $\alsfivepi^5$ term of
$\Delta_{gg}^{\mb=0}$ has the same order of magnitude as the $\alsfivepi^4$ term of
$\Delta_{\rm light}$.  Note that the latter is only about twice as large as
the $\alsfivepi^4$ contribution to $\Delta_{\rm top}$, obtained from the sum of
the three numbers in the last line of Eq.~(\ref{eq::Delta_top}); this amounts
to $0.0006593$.

It is a disturbing feature of $\Delta_{\rm light}$ that the $\alsfivepi^3$ and
$\alsfivepi^4$ terms deviate by less than 30\%. Furthermore they have opposite
signs. Therefore, it is interesting to add $\Delta_{\rm light}$ and
$\Delta_{\rm top}$ which leads to
\begin{eqnarray}
  1 + \Delta_{\rm light} + \Delta_{\rm top}
  &\approx&
  1 + 0.2033 + 0.04208 + 0.004490 - 0.0007090
  \,,
  \label{eq::Delta_lighttop}
\end{eqnarray}
where the different loop orders are kept separate.  We observe a
reduction by a factor of about six between the three- and four-loop contributions;
the convergence of the sum is significantly better than that of the individual expressions.

\begin{figure}[t]
  \centering
  \includegraphics[width=.9\textwidth]{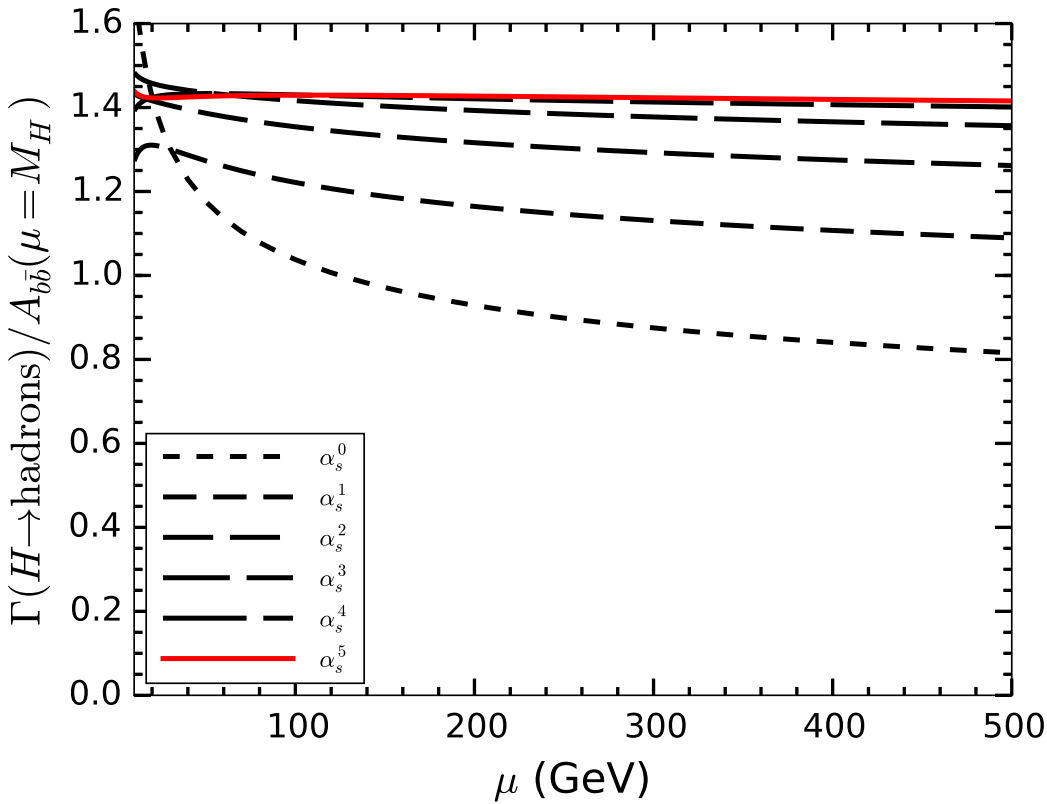}
  \caption{\label{fig::gam_mu}$\Gamma(H\to\mbox{hadrons})/A_{b\bar{b}}(\mu=\MH)$
    as a function of the renormalization scale $\mu$.}
\end{figure}

Finally we show, in Fig.~\ref{fig::gam_mu}, the dependence of $\Gamma(H\to
\mbox{hadrons})$ on the renormalization scale $\mu$.  We plot
$\Gamma(H\to\mbox{hadrons})/A_{b\bar{b}}(\mu=\MH)$, which means that for the
leading order (short-dashed) curve we have
$\Gamma(H\to\mbox{hadrons})/A_{b\bar{b}}(\mu=\MH)=1$ for $\mu=\MH$. The six
curves represent (from bottom to top, i.e. from the short-dashed to the solid
curve) the predictions of order $\als^0$, \ldots, $\als^5$, where
$\als^5$ terms are only included for $\Delta_{gg}^{\mb=0}$.  $\mu$ is
varied between 10~GeV and 500~GeV which is significantly larger than the usual
range spanned between $\MH/2$ and $2\MH$. 
Nevertheless, one observes a steady
flattening of the curves when including higher order corrections; the
result represented by solid line is almost
$\mu$-independent.


\section{\label{sec::concl}Conclusions}

We complete the corrections of order $\als^4$ to the hadronic
decay rate of the Standard Model Higgs boson by computing the
top quark--induced contributions in an effective field-theory framework. This
requires the calculation of four-loop propagator-type integrals.
Our new corrections are numerically of the same order of magnitude as
the purely massless contribution~\cite{Baikov:2005rw}, however they have
an opposite sign. We provide all analytic results presented
in this paper in a computer-readable format~\cite{progdata},
making it straightforward to implement the corrections in existing
computer codes which evaluate decay rates of the Higgs boson.
Finally, we want to mention that
$\Gamma(H\to\mbox{hadrons})$ is one of very few physical quantities
for which five terms of the perturbative expansion are known and the
perturbative expansion can be studied, see Eq.~(\ref{eq::Delta_lighttop}).



\section*{Acknowledgements}

We thank Konstantin Chetyrkin for providing analytic results for $\Delta_{11}$
and $\Delta_{22}$ containing the full $\mu$ dependence and Vladimir
Smirnov for analytical results for the four-loop master integrals.  This
work is supported by the BMBF grant 05H15VKCCA.



\begin{appendix}

\section{\label{app::Del12}$\Delta_{11}$ and $\Delta_{12}$ in terms of Casimir colour factors}

In terms of the Casimir invariants of SU($N$), $\Delta_{11}$ is given by
\begin{flalign}
\Delta_{11} =
1 & + \alsfivepi^{\hphantom{2}}\,
\bigg[
	\ca  \,  \Big(
		\frct{73}{12}
		+ \frct{11}{6} \, \LH
		\Big)
		- \nl  \,  \Big(
		\frct{7}{6}
		+ \frct{1}{3} \, \LH
	\Big)
\bigg]
\nonumber\\&
+ \alsfivepi^2 \, \bigg[
	\catwo  \,  \Big(
		\frct{37631}{864}
		- \frct{121}{24} \, \zetatwo
		- \frct{55}{8} \, \zetathr
		+ \frct{313}{16} \, \LH
		+ \frct{121}{48} \, \LHtwo
	\Big)
\nonumber\\& \hphantom{{} + \alsfivepi^2 \, \bigg[}
	+ \nl \, \cf  \,  \Big(
		- \frct{131}{48}
		+ \frct{3}{2} \, \zetathr
		- \frct{1}{2} \, \LH
	\Big)
\nonumber\\& \hphantom{{} + \alsfivepi^2 \, \bigg[}
	+ \nl \, \ca  \,  \Big(
		- \frct{6665}{432}
		+ \frct{11}{6} \, \zetatwo
		- \frct{1}{4} \, \zetathr
		- \frct{85}{12} \, \LH
		- \frct{11}{12} \, \LHtwo
	\Big)
\nonumber\\& \hphantom{{} + \alsfivepi^2 \, \bigg[}
	+ \nltwo  \,  \Big(
		\frct{127}{108}
		- \frct{1}{6} \, \zetatwo
		+ \frct{7}{12} \, \LH
		+ \frct{1}{12} \, \LHtwo
	\Big)
\bigg]
\nonumber\\&
+ \mbmh \, \Bigg\{
	6 \, \alsfivepi
	+ \alsfivepi^2 \, \bigg[
		\ca  \,  \Big(
			55
			+ 6 \, \zetathr
			+ \frct{33}{2} \, \LH
		\Big)
\nonumber\\&
		\hspace{7mm}{} + \cf \,  \Big(
			\frct{101}{2}
			- \frct{9}{2} \, \zetatwo
			- 9 \, \zetathr
			+ \frct{105}{4} \, \LH
			+ \frct{9}{4} \, \LHtwo
		\Big)
		- \nl  \,  \Big(
			\frct{15}{2}
			+ 3 \, \LH
		\Big)
	\bigg]
\Bigg\}
+ {\cal O}\left( \alsfivepi^3 \right),&
\end{flalign}
where $C_A=N$ and $C_F=(N^2-1)/(2N)$.
$\Delta_{12}$ reads
\begin{flalign}
\Delta_{12} =
& + \alsfivepi^{\hphantom{2}}\,
\bigg[
	\cf \,  \Big(
		- 23
		- 6 \, \LH
	\Big)
\bigg]
\nonumber\\
& + \alsfivepi^2 \, \bigg[
	\cftwo  \,  \Big(
		- \frct{907}{8}
		+ 18 \, \zetatwo
		+ 18 \, \zetathr
		- \frct{123}{2} \, \LH
		- 9 \, \LH^2
	\Big)
\nonumber\\& \hphantom{{} + \alsfivepi^2 \, \bigg[}
	+ \ca \, \cf  \,  \Big(
		- \frct{3815}{24}
		+ 11 \, \zetatwo
		+ 31 \, \zetathr
		- \frct{175}{3} \, \LH
		- \frct{11}{2} \, \LH^2
	\Big)
\nonumber\\& \hphantom{{} + \alsfivepi^2 \, \bigg[}
	+ \nl \, \cf  \,  \Big(
		\frct{283}{12}
		- 2 \, \zetatwo
		- 4 \, \zetathr
		+ \frct{28}{3} \, \LH
		+ \LH^2
	\Big)
\bigg]
\nonumber\\
& + \alsfivepi^3 \, \bigg[
	\cfthr  \,  \Big(
		- \frct{29545}{64}
		+ 135 \, \zetatwo
		+ \frct{663}{4} \, \zetathr
		- \frct{135}{2} \, \zetafiv
\nonumber\\&\hphantom{{} + \alsfivepi^3 \, \bigg[ \cfthr  \,  \Big(}
		- \Big[\frct{8631}{32} - \frct{81}{2} \, \zetatwo
			- \frct{81}{2} \, \zetathr\Big] \, \LH
		- \frct{135}{2} \, \LHtwo
		- \frct{27}{4} \, \LHthr
	\Big)
\nonumber\\&\hphantom{{} + \alsfivepi^3 \, \bigg[}
	+ \ca \, \cftwo  \,  \Big(
		- \frct{108241}{96}
		+ \frct{657}{2} \, \zetatwo
		+ \frct{3189}{8} \, \zetathr
		- \frct{435}{4} \, \zetafiv
\nonumber\\&\hphantom{{} + \alsfivepi^3 \, \bigg[+ \ca \, \cftwo  \,  \Big(}
		- \Big[\frct{23585}{32} - \frct{297}{4} \, \zetatwo
			- \frct{477}{4} \, \zetathr\Big] \, \LH
		- \frct{657}{4} \, \LHtwo
		- \frct{99}{8} \, \LHthr
	\Big)
\nonumber\\&\hphantom{{} + \alsfivepi^3 \, \bigg[}
	+ \catwo \, \cf  \,  \Big(
		- \frct{5886949}{5184}
		+ \frct{1039}{6} \, \zetatwo
		+ \frct{3187}{8} \, \zetathr
		- \frct{25}{4} \, \zetafiv
\nonumber\\&\hphantom{{} + \alsfivepi^3 \, \bigg[+ \catwo \, \cf  \,  \Big(}
		- \Big[\frct{18923}{36} - \frct{121}{4} \, \zetatwo
			- \frct{341}{4} \, \zetathr\Big] \, \LH
		- \frct{1039}{12} \, \LHtwo
		- \frct{121}{24} \, \LHthr
	\Big)
\nonumber\\&\hphantom{{} + \alsfivepi^3 \, \bigg[}
	+ \nl \, \cftwo  \,  \Big(
		\frct{5803}{24}
		- \frct{225}{4} \, \zetatwo
		- \frct{207}{2} \, \zetathr
		- \frct{9}{2} \, \zetafou
		+ 30 \, \zetafiv
\nonumber\\&\hphantom{{} + \alsfivepi^3 \, \bigg[+ \nl \, \cftwo  \,  \Big(}
		+ \Big[\frct{1067}{8} - \frct{27}{2} \, \zetatwo - 27 \, \zetathr\Big] \, \LH
		+ \frct{225}{8} \, \LHtwo
		+ \frct{9}{4} \, \LHthr
	\Big)
\nonumber\\&\hphantom{{} + \alsfivepi^3 \, \bigg[}
	+ \nl \, \ca \, \cf  \,  \Big(
		\frct{209815}{648}
		- \frct{703}{12} \, \zetatwo
		- \frct{131}{2} \, \zetathr
		+ \frct{9}{2} \, \zetafou
		- 5 \, \zetafiv
\nonumber\\&\hphantom{{} + \alsfivepi^3 \, \bigg[+ \nl \, \ca \, \cf  \,  \Big(}
		+ \Big[\frct{11705}{72} - 11 \, \zetatwo
			- \frct{35}{2} \, \zetathr\Big] \, \LH
		+ \frct{703}{24} \, \LHtwo
		+ \frct{11}{6} \, \LHthr
	\Big)
\nonumber\\&\hphantom{{} + \alsfivepi^3 \, \bigg[}
	+ \nltwo \, \cf  \,  \Big(
		- \frct{25627}{1296}
		+ \frct{14}{3} \, \zetatwo
		+ 5 \, \zetathr
\nonumber\\&\hphantom{{} + \alsfivepi^3 \, \bigg[+ \nltwo \, \cf  \,  \Big(}
		- \Big[\frct{407}{36} - \zetatwo - 2 \, \zetathr\Big] \, \LH
		- \frct{7}{3} \, \LHtwo
		- \frct{1}{6} \, \LHthr
	\Big)
\bigg]
+ {\cal O}\left( \alsfivepi^4 \right)\,.&
\end{flalign}

\end{appendix}


\end{document}